\documentclass[aps,pre,twocolumn,10pt,amsmath,amssymb,showpacs]{revtex4-1}
\usepackage{graphicx}
\usepackage{dcolumn}
\usepackage{bm}
\usepackage[centerlast]{caption}
\usepackage{subcaption}
  
\def\be{\begin{equation}}
\def\ee{\end{equation}}

\def\bea{\begin{eqnarray}}
\def\eea{\end{eqnarray}}

\begin{document}
\title{Critical Casimir forces along the iso-fields}

\author{M.~Zubaszewska$^{1}$, A.~Macio\l ek$^{2,3,4}$, A.~Drzewi\'{n}ski$^{1}$}
\affiliation{$^{1}$Institute of Physics, University of Zielona G\'{o}ra, \\
ul. Prof. Z. Szafrana 4a, 65-516 Zielona G\'{o}ra, Poland}
\affiliation{$^{2}$Max-Planck-Institut f{\"u}r Intelligente Systeme, Heisenbergstr.~3,
D-70569 Stuttgart, Germany}
\affiliation{$^{3}$IV Institut f\"ur Theoretische  Physik,
Universit\"at Stuttgart, Pfaffenwaldring 57, D-70569 Stuttgart, Germany }
\affiliation{$^{4}$Institute of Physical Chemistry, Polish Academy of Sciences,
Kasprzaka 44/52, PL-01-224 Warsaw, Poland}

\date{\today}

\begin{abstract}
Using quasi-exact numerical density-matrix renormalization-group techniques we calculate the critical Casimir
force  for a two-dimensional ($2D$) Ising strip with equal strong surface fields, along the thermodynamic paths
corresponding to the fixed nonzero bulk field $h\ne 0$. Using the Derjaguin approximation we also determine
the critical Casimir force and its potential for two discs.  We find that varying the temperature along
the iso-fields lying between the bulk coexistence and the capillary condensation critical point leads to
a dramatic increase of the critical Casimir interactions with a qualitatively different functional dependence
on the temperature than along $h=0$. These findings  might be  of relevance for biomembranes, whose heterogeneity
is recently interpreted as being connected with a critical behavior belonging to the $2D$ Ising universality class.
\end{abstract}


\maketitle

\section{Introduction}

Effective forces  arising  between two surfaces 
confining a fluid close to its critical point, known in the literature as  critical Casimir forces $F_C$
\cite{FdG},
  have been studied theoretically and experimentally
for systems belonging to various bulk and surface universality classes of critical phenomena.
Different geometries of confining surfaces have also
 been considered \cite{krech:99:0,dantchev,gambassi}.
These diverse studies show that critical Casimir forces 
are very sensitive to tiny changes in temperature:
a slight variation in the temperature can lead to  pronounced changes 
in the range and  magnitude of $F_C$ - even the sign of the  force reverses. Such sensitivity is rather rare but very valuable
feature of effective interactions.
It might be of   importance for systems such as 
  colloidal  suspensions with   a near-critical solvent. There, by tuning 
the critical Casimir interactions arising between colloidal particles one can  control   the phase behavior  
\cite{bechin,thomas1,schall,schall1} of colloids, their  aggregation and the stability of colloidal  
suspensions \cite{schall2,thomas2}. 

For colloids immersed in a binary liquid mixture, it has been observed  that 
 by varying temperature towards  the bulk demixing phase boundary
 from a one-phase region of a solvent at its constant composition, the colloidal particles
  coagulate reversible   at  well defined near-critical temperatures
\cite{Beysens-et:1985,Beysens-et:1999,Bonn-et:2009,Buzzaccaro-et:2010}. 
The strongest coagulation of colloids is observed for off-critical compositions of the solvent, i.e., 
at thermodynamic states of the solvent
which correspond to a nonzero  conjugate ordering field $h$ of the order
parameter of the solvent. Recall that $h$ is the deviation $\mu_A -\mu_B -(\mu_A - \mu_B)_c$  of the chemical potential 
difference of the two species $A$ and $B$ forming a  binary liquid mixture from its critical value,
or  a chemical potential deviation $\mu - \mu_c$ from its critical value  in the case of a simple fluid.

\bigskip
\begin{figure}[tbh]
\centering
\includegraphics[scale=0.3,angle=0]{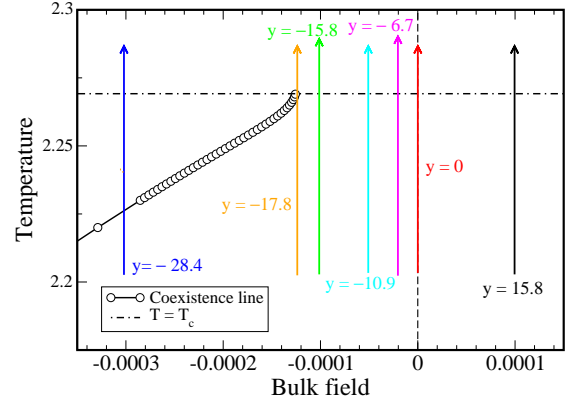}
\caption{(Color online) Symbols connected by a solid line show the  line of zero total magnetization for the Ising strip
of the width $L=501$ and the surface field $h_1=6.32$ corresponding to the scaling variable $z=20 000$.
Below the pseudo-critical point of the strip $(T_{c,L},h_{c,L})$ (not determined here), this line can be
identified with the pseudo-coexistence capillary condensation. Thermodynamic paths of fixed scaling variable
$y=L|h|^{8/15}/\xi_h^0$ (as indicated on the plot) along which we have calculated the critical Casimir force
are shown by solid lines. For $L=501$ the chosen values of $y$ correspond to the bulk fields (subsequently
from the left to the right): $|h|=0.003, 0.000123, 0.0001, 0.00005, 0.00002$.}
\label{fig:01}
\end{figure}

Various theoretical studies for systems belonging to the Ising universality class
provide some evidence \cite{Buzzaccaro-et:2010,dme1,dme2,colloids1b,onuki,thesis} 
for a pronounced dependence of critical Casimir forces $F_C$  on $h$.
The only geometry considered so far in  this context is the  geometry of two parallel plates, however, 
the critical Casimir scaling functions for
parallel plates can be  used to approximate the ones for the sphere-plate or sphere-sphere geometry. This can be done  
by using the Derjaguin approximation in the limit of  surface-to-surface
separations  small in comparison with  the sphere radius \cite{Derjaguin:1934}.
Although in most experimental realizations,  the temperature is varied at the fixed value of $h$, 
no results for  $F_C$ along such thermodynamic paths are available. 
In the present paper we report a systematic study of the critical  Casimir forces along several bulk iso-fields
 for  a $2D$ Ising model with a strip geometry.
We assume that  boundaries of the strip are equal and  belong to the so-called
 normal transition surface universality class  $(+,+)$ \cite{diehl}, which is characterized by a strong effective surface field acting on the 
corresponding order parameter of a system. Normal transition surface universality class is an appropriate characterization
of  a critical fluid in a presence of an external wall or a substrate \cite{diehl1}.
In the case of colloidal suspensions immersed in a binary liquid mixture,
 surface fields describe a preference of colloidal particles  for one of the two components of 
a binary liquid mixture.  
 We consider the bulk iso-fields  which lie on both sides of the  bulk coexistence
$h=0$ and close to it, where we expect the most pronounced variation of the critical Casimir force.
On that side of the bulk coexistence line which corresponds to the bulk phase  not preferred by the 
surfaces (e.g., a negatively magnetized phase
for positive surface fields), 
the bulk iso-fields cross the  pseudo-coexistence line of capillary condensation, i.e., the
bulk pseudo-coexistence line shifted away from $h=0$ due to the confinement.
We explore in some detail the behavior of $F_C$ as a function of temperature
in the region between the capillary condensation line  and  the bulk coexistence line.

In order to calculate the free energy of a system we use  a quasi-exact numerical 
density-matrix renormalization-group method (DMRG).
The DMRG method \cite{DMRG}, which is based on the transfer matrix approach \cite{Nishino}, provides
a numerically very efficient iterative truncation algorithm for constructing the effective transfer matrices for
strips of fixed width and infinite length \cite{AD}. At present, strips of widths up to $L=700$ lattice constants
can be studied. The quantitative data for thermodynamic quantities, especially the free energy, are very accurate
and can be provided in a wide range of temperatures and in the presence of arbitrary bulk and surface fields.
This allows to determine the scaling functions of the critical Casimir forces.

It is know that the critical Casimir scaling function  depends in a nontrivial way  on the spatial dimension. 
However, the existing results
along other thermodynamic paths, e.g., along isotherms or along the bulk coexistence, show that 
the qualitative behavior of $F_C$ 
in higher spatial dimensions is similar to the one in $2D$; we expect the same to be true also along  the bulk iso-fields.
Results of our study are directly applicable to the lipid  membranes. These are 
$2D$ liquids  consisting of two (or more) components, such as cholesterol 
and the saturated and unsaturated lipid, which   undergo separation into two liquid phases, one rich in the 
first two components and the other rich in the third \cite{membranes1}.
Lipid membranes serve as  model systems for 
cell plasma membranes \cite{LS}.  
Recent experiments suggests that cell membranes 
are tuned to the miscibility critical point of $2D$ Ising model \cite{membranes}
so that the critical Casimir forces may arise between
macromolecules embedded in the membrane \cite{sehtna}.

\bigskip
\begin{figure}[htb]
\centering
\includegraphics[scale=0.3,angle=0]{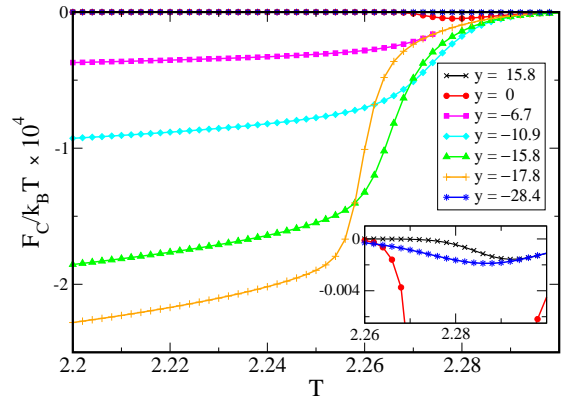}
\caption{(Color online) The critical Casimir forces for the Ising strip with the  width  $L=301$ and the surface field
$h_1=8.15$ providing the scaling parameter $z=20 000$. The values of the bulk field ensure that the scaling
field $y$ for which the curves have been calculated, is the same as in Fig.~(\ref{fig:01}). The inset shows
the vicinity of the bulk critical point, where only minima are present. The solid lines are guides for eyes.}
\label{fig:02}
\end{figure}

\section{The model}
\label{sec:1}

We consider a two-dimensional  Ising model on the  square
lattice $M \times L$ with a strip geometry and with cyclic boundary conditions in $(1,0)$ direction.
The energy for a configuration $\{ s \}$ of spins is given by the Hamiltonian:
\[
{\cal H} = -J \left( \sum_{<ij,i'j'>} s_{i,j} s_{i',j'} - h_1 \sum\limits_{{\rm surface}\atop{\rm spins}} s_{i,j}
- h \sum\limits_{{\rm all}\atop{\rm spins}}  s_{i,j} \right),
\]
with $J>0$ and $s_{i,j} = \pm 1$, where $(i,j)$ labels the site of the lattice. The first sum is over nearest neighbours, while the second sum is performed
over  spins at the both surfaces. We are interested in the limit  $M\to \infty$  with finite $L$.
Bulk and surface fields are measured in the units of the coupling constant $J$ and the distances are measured in units of the lattice constant $a$.

In the two-dimensional Ising model, the  scaling fields describing the deviation from the bulk criticality are
the reduced temperature $\tau = (T-T_c)/T_c$, where $T_c\simeq 2.26919$ is the bulk critical temperature, and the bulk magnetic field $h$. The presence of a wall introduces
an additional field: the surface field $h_1$. 
For a strip of finite width $L$, the scaling variables  related to the above scaling fields are $x=\mathrm{sgn}(\tau) L/\xi_{\tau}= L \tau/\xi_0^+$, $y=\mathrm{sgn}(h) L/\xi_{h}=L|h|^{8/15}/\xi_{h}^{0}$, 
and
$z=L/l_{h_1}=L|h_1|^2$.
For $2D$ Ising model  $\xi_{0}^{+} = 0.5673$ and  $\xi_{h}^{0} = 0.233 \pm 0.001$ \cite{TF,DMC}. Here, 
$\xi_{\tau}:=\xi(\tau,h=0)=\xi_{0}^+|\tau|^{-\nu}$,  $\xi_{h}:=\xi(\tau=0,h)=\xi_h^0|h|^{-\nu/(\beta\delta)}$, where $\xi(\tau,h)$
is the true bulk correlation length that governs the exponential decay of the bulk order parameter correlation functions.
The length scale associated with the surface field $l_{1}=l_1^0l_{h_1}=l_1^0|h_1|^{-\nu/\Delta_1^{ord}}$, where the amplitude
$l_1^0 \simeq 2.8\xi_0^+$ for the  $2D$ Ising model \cite{vmd}.
 $\nu=1, \beta=1/8$ and $\delta=15$ are   $2D$ Ising model values of standard bulk critical exponents and $\Delta^{ord}_1=1/2$
 is the surface counterpart of bulk gap exponent $\Delta$.
Because the scaling variable corresponding to a surface field is set to  $z=20000$, the effect of a surface magnetic field
reaches saturation, which corresponds to the fixed point $|h_1|=\infty$ of the normal transition \cite{diehl}. 
Moreover, in such a case the pseudo-coexistence line  in the strip with fixed width is maximally moved away from
the temperature axis. The chosen thermodynamic paths are the most informative for
the present system. They are shown together with the pseudo-coexistence line $h_{co}(T)$  in  Fig.~(\ref{fig:01}).
$h_{co}(T)$ have been identified
as those positions $(h, T)$ (below $T_c$) in the phase diagram, where the total magnetization of the strip
vanishes, i.e., $\sum_{j=1}^L m_j=0$, with $ m_j = \langle s_{i,j} \rangle$.  Because non-analytic behavior is
rounded in $2D$, it is not possible to uniquely determine  the pseudo-critical point \cite{dme1}.

\bigskip
\begin{figure}[htb]
\centering
\includegraphics[scale=0.3,angle=0]{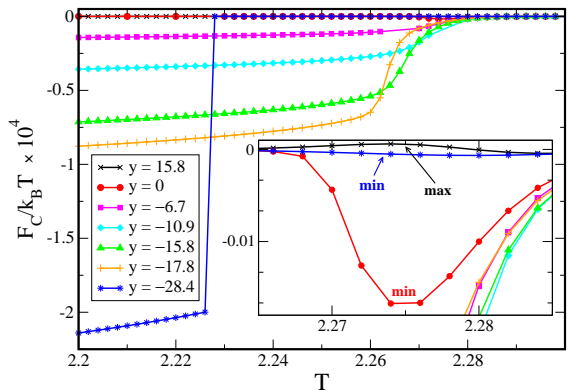}
\caption{(Color online) The critical Casimir forces for $L=501$ and the surface field $h_1=6.32$ providing the scaling
parameter $z=20 000$. The inset presents the vicinity of the bulk critical point. The solid lines are
guides for eyes.}
\label{fig:03}
\end{figure}

\section{Results}
\label{sec:2}

\subsection{Film geometry}
\label{subsec:film}

In order to find the critical Casimir force, we first calculate the excess free
energy per unit length in $(1,0)$ direction
$f^{\rm ex} (L) \equiv (f(L) - f_b ) L$, where $f_b$ is the bulk free energy per spin. For the vanishing
bulk field, $f_b$ is known from an analytical exact  solution of the two-dimensional Ising model \cite{Onsager}. 
For non-vanishing
bulk fields, we have to calculated the bulk  free energy numerically.
We have done that in two steps, first, we have determined the free energies for several  strips of widths 
up to 600 lattice constants  with free boundaries ($h_1=0$)  by using the DMRG method. 
 Next, the data have been  extrapolated to infinity by
the Bulirsch-Stoer method \cite{BS}.
A precise determination of the bulk free energy along the chosen thermodynamic paths 
is very costly numerically, but the high  accuracy  data for $f_b$ are  
 absolutely  necessary in order to obtained a reliable
results for the critical Casimir forces.

\bigskip
\begin{figure}[htb]
\centering
\includegraphics[scale=0.3, angle = 0]{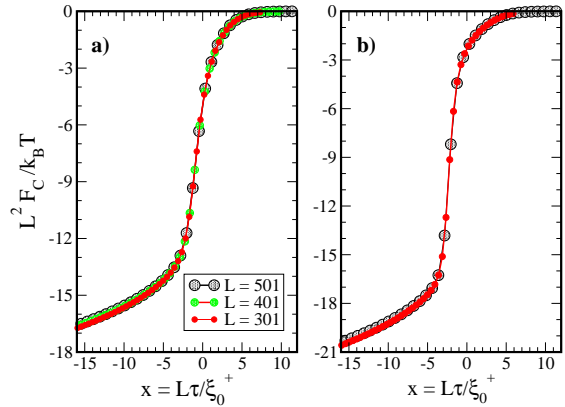}
\caption{(Color online) The scaling function of the critical  Casimir force as a function of the scaling variable  $x=L\tau/\xi_0^+$
at $z=20 000$ and for  (a) $y = - 15.8$, and (b) $y = -17.8$. The solid lines are guides for eyes.}
\label{fig:04}
\end{figure}

The critical Casimir force for our system is defined as
\begin{equation}
 F_C = - (\partial f^{\rm ex} (L) / \partial L)_{h,T}.
\end{equation}
In our calculations we approximate the above derivative by a finite difference
\begin{equation}
 F_C = - (1/2)\left[f^{\rm ex} (L+2) - f^{\rm ex} (L))\right].
\end{equation}

From the finite-size scaling theory \cite{FSS} and the renormalization-group analysis  for  the film geometry  it follows  that the critical Casimir force $F_C$ takes the following scaling form 
\begin{equation}
 \label{eq:scal}
\frac{F_C(T,L,h,h_1)}{k_BT}=\frac{1}{L^d}\vartheta(x,y,z).
\end{equation}
$d=2$ in two dimensions and $\vartheta(x,y,z)$ is the universal scaling function, where the scaling variables $x, y$
and $z$ were defined in the previous section.

As one can see in Fig.~(\ref{fig:02}), the curves representing the critical Casimir force  for the strip of
the width $L=301$ behave smoothly as a function
of temperature for all  values of $y$. It means that the accuracy of the DMRG calculation
is high enough for this width of the strip.
In Fig.~(\ref{fig:03}), the similar results but for a wider strip with $L=501$ are shown.

\begin{figure}[t]
\centering
\includegraphics[scale=0.3, angle=0]{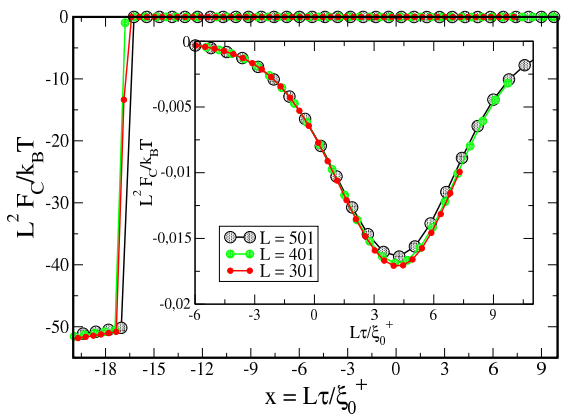}
\caption{(Color online) The scaling function of the critical Casimir force as a function of the scaling variable  $x=L\tau/\xi_0^+$
at $z=20 000$ and $y= -28.4$. The inset presents the near-critical behavior. The solid lines are guides for eyes.}
\label{fig:05}
\end{figure}

Along the line $y = 0$, the negative critical Casimir force exhibits a minimum  above the bulk critical temperature
(see the curve marked by full circles in Figs.~(\ref{fig:02}) and (\ref{fig:03})).
This result is already known from the analytical calculations by Evans and Stecki \cite{Bob}.
Upon decreasing the temperature along the  lines located  between the bulk coexistence $y=0$ and
the pseudo-coexistence line $h_{co}(T)$, i.e., along the lines with  $y= -6.7, -10.9, -15.8,$ and $-17.8$,
a rapid drop of the critical Casimir force occurs at a rather narrow interval of temperatures near $T_c$
followed by a more gradual, nearly linear decay towards more negative values. The further from the axis $y=0$,
the more rapid is the initial drop. When the thermodynamic path intersects the pseudo-coexistence curve,
the decline begins to resemble a sudden jump (see the curve denoted by stars corresponding to $y=-28.4$
in Fig.~(\ref{fig:03}) for $L=501$). For thermodynamic paths that cross the pseudo-coexistence line, we observe
in addition to the jump of the critical Casimir force at the crossing temperature, also the minimum located close
to the bulk critical point above $T_c$. It can clearly be seen for the strip with  $L=501$ at $y = - 28.4$ where
the thermodynamic path is the most shifted from the bulk coexistence line (the line with stars in Fig.~(\ref{fig:03})
and in its inset). For the positive values of the bulk field (identical to the sign of the surface field),
the critical Casimir force seems to behave uniformly. It is very small (negative) and only close to the bulk
critical point a weak minimum can be observed above $T_c$. The maximum occurring for $L=501$ is not physical;
it is an artefact of the inaccuracy in determining the bulk free energy.

\begin{figure}[t]
\centering
\includegraphics[scale=0.34, angle=0]{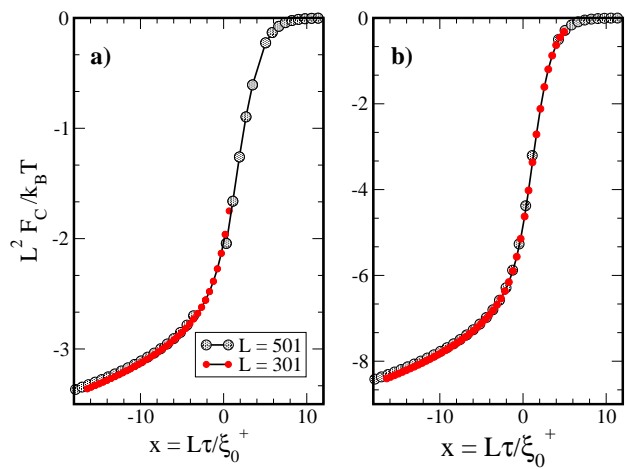}
\caption{(Color online) The scaling function of the critical  Casimir force as a function of the scaling variable  $x=L\tau/\xi_0^+$
at $z=20 000$ and for  (a) $y = - 6.7$, and (b) $y = -10.9$. }
\label{fig:06}
\end{figure}

In Figs~(\ref{fig:04})(a) and (b) and (\ref{fig:05})  we plot  the critical  Casimir force calculated for
three different widths of the strip: $L=301, 401$ and $501$  and scaled by $L^2$ as a function of the scaling
variable $x=L\tau/\xi_0^+$ for  fixed values of $y=  -15.8, -17.8$  and $-28.4$, respectively.
We find a very good data collapse, which confirms that the scaling along the iso-fields  holds even 
for $y=-28.4$, which is relatively far  from the bulk coexistence.

\begin{figure}[htb]
\centering
\includegraphics[scale=0.3, angle=0]{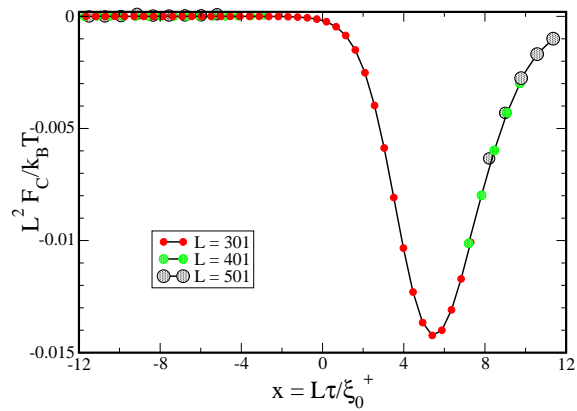}
\caption{(Color online) The scaling function of the critical Casimir force as a function of the scaling variable  $x=L\tau/\xi_0^+$
at $z=20 000$ and $y = 15.8$. }
\label{fig:07}
\end{figure}

At the values of $y$ closer to the bulk coexistence, results for $L=501$ and $401$ deviate  from the master curve 
in the region  very close to the critical point  where our DMRG-based extrapolation  fails.
These  deviations are much more pronounced for positive value of $y= 15.8$,  where  the  magnitude of the Casimir
force is  very small and lead to the occurrence of the  unphysical maximum and the positive values of the Casimir
scaling function near $x=0$.  In Figs~(\ref{fig:06})(a) and (b) and (\ref{fig:07}) we show the Casimir scaling
functions for $y =  -6.7, -10.9$, and $y = 15.8$, respectively. Data which deviate from the common function are
not shown.

\bigskip
\bigskip

\subsection{Disc-disc geometry}
\label{subsec:discs}

As already mentioned in  Introduction, fluctuations of the 
membrane concentration can lead to the critical Casimir interactions between various  membrane inclusions. 
Modeling the latter by  discs and employing  the Derjaguin approximation,  we can use our results  in order to calculate
the critical Casimir interaction and its potential for two identical discs embedded in the $2D$ Ising model.
Results available for the strongly adsorbing identical discs ($(+,+)$ surface universality class) are very limited.
The analysis based on the conformal invariance is restricted to 
 the bulk critical point ( $T=T_c$, $h=0$ ). It predicts  that in the so called protein limit of small discs far apart, i.e., $\xi \gg L_0 \gg R$,
 where $L_0$ is the surface-to-surface separation and $R$ is the radius of a disc, the critical Casimir potential is given by \cite{BE,MS}
 \begin{equation}
  U_C^{(dd)}(L)/k_BT_C = - \sqrt {2} \left(\frac{R}{L}\right)^{1/4}.
 \end{equation}
In the opposite Derjaguin limit one has
\begin{equation}
\label{eq:confinv}
  U_C^{(dd)}(L)/k_BT_C = -\frac{\pi^2}{24} \left(\frac{R}{L}\right)^{1/2}.
 \end{equation}
At the bulk critical point and for three  temperatures above $T_c$, the Casimir potential for two disc-like objects
embedded in the $2D$  Ising lattice has been determined from the Monte Carlo simulations. 
However, no results for the Casimir potential in the full range of temperatures or, more generally,
for the scaling functions are available.
 Therefore, even approximate results as those obtained using the Derjaguin approximation
are valuable. Various studies \cite{Hanke} 
show that within the Derjaguin approximation   qualitative features of the scaling functions are correct.

\begin{figure}[htb]
\centering
\includegraphics[scale=0.62]{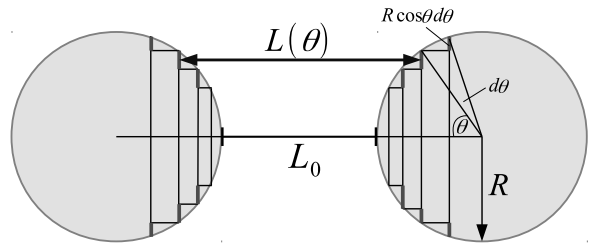}
\caption{Geometry of the Derjaguin approximation for the disc-disc geometry.}
\label{fig:08}
\end{figure}

The Derjaguin approximation is valid if the radius $R$ of discs is much larger than the minimal separation $L_0$
between their surfaces.
In the  spirit of the Derjaguin approximation in two dimensions, the curved edge of a  disc is considered to be made up of 
infinitely thin stairs (steps) with the  length $dS(\theta)$. These thin stairs are parallel to each other 
for two discs and are at a normal distance $L(\theta)=L_0+2R(1-\cos\theta)$ from an opposing identical stair (see Fig.~\ref{fig:08}).
 Using the assumption of additivity of the forces  underlying the Derjaguin approximation, we assume that the contribution $dF_C^{(dd)}$
 of each single pair of stairs 
to the total critical Casimir force $F_C^{(dd)}$ is given by the sum of the critical Casimir forces which would act
 in the  film geometry on portions of $dS(\theta)=R(\theta+d\theta)-R(\theta)\simeq R\cos\theta d\theta$ .

According to eq. (\ref{eq:scal}) this leads to the following expression for the force acting on a single stair:
\begin{equation}
\label{eq:diff}
 \frac{dF_C^{(dd)}}{k_BT}=\frac{dS(\theta)}{L^2(\theta)}\vartheta(x(\theta),y(\theta),z(\theta)),
\end{equation}
where $\vartheta$ is the scaling function of the critical Casimir force in the film geometry.
Assuming that the scaling variable $z$ is fixed at the value corresponding to the $(+,+)$ boundary conditions,
the above equation can be written as

\begin{equation}
 \frac{dF_C^{(dd)}}{k_BT}=\frac{dS(\theta)}{L^2(\theta)}\vartheta(x(\theta),y(\theta)).
\end{equation}
\begin{figure}
        \centering
        \begin{subfigure}[b]{0.4\textwidth}
                \centering
                \includegraphics[scale=0.3]{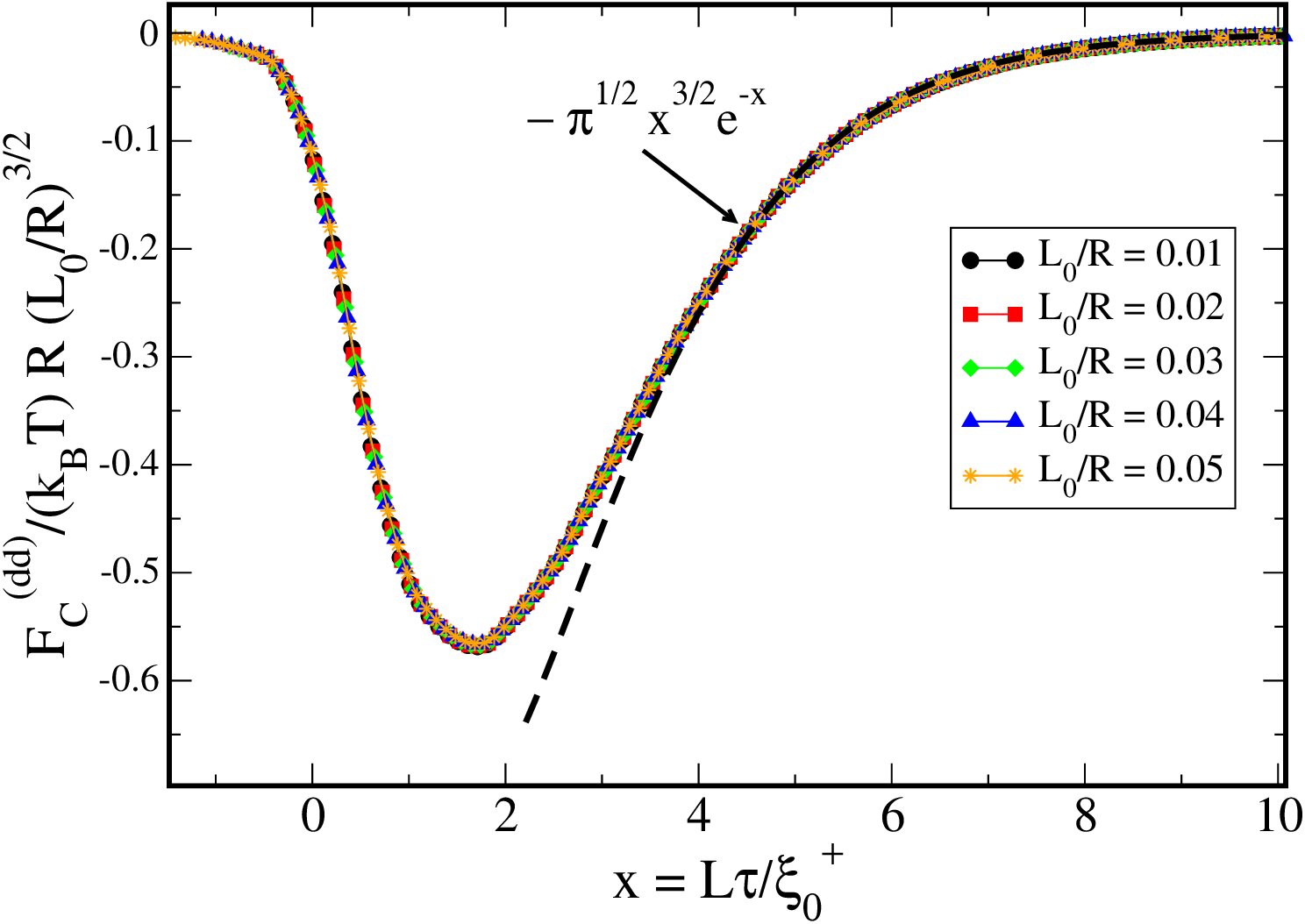}
                \caption{} 
        \end{subfigure}%
        
         \vspace{0.85cm}
         
        \begin{subfigure}[b]{0.4\textwidth}
                \centering
                \includegraphics[scale=0.3]{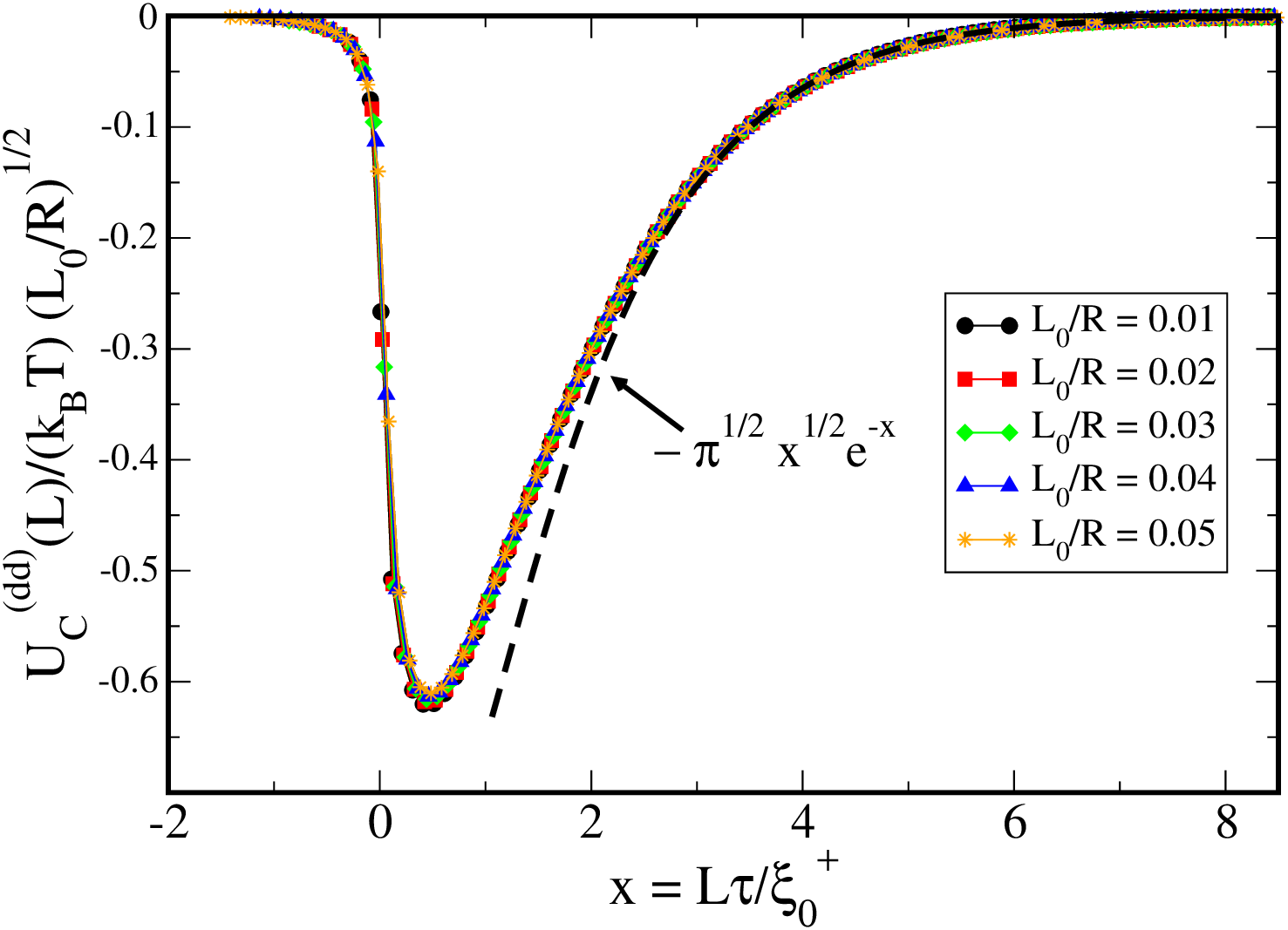}
                \caption{}
    
        \end{subfigure}
\caption{(Color online) The rescaled  critical Casimir force $F_C^{(dd)}/(k_BT)(L_0/R)^{3/2}*R$ (a) and its potential
$U_C^{(dd)}/(k_BT)(L_0/R)^{1/2}$ (b) the for two identical discs with radii $R=300, 500$ and 1000 as
a function of the scaling variable  $x=L\tau/\xi_0^+$ at $z=20 000$ and $y = 0$, i.e., along the bulk
coexistence. Data calculated for various ratios $R/L_0$ form a common curve. Dashed line corresponds to
$F_C^{(dd)}/(k_BT)$ and $U_C^{(dd)}/(k_BT)$ calculated using the asymptotic behavior of the critical
Casimir force for a $(+,+)$  strip  (see the main text).}
\label{fig:09}
\end{figure}
The total force $F_C^{(dd)}$ is obtained by summing all the contributions $dF_C^{(dd)}(\theta)$ from the stairs up 
to the maximal angle $\theta_M=\frac{\pi}{2}$.
\begin{equation}
\label{total}
\frac{F_C^{(dd)}}{k_BT}=\int_{\theta=0}^{\theta_M=\frac{\pi}{2}} \frac{dS(\theta)}{L^2(\theta)}\vartheta(x(\theta),y(\theta)).
\end{equation}
\begin{figure}
        \centering
        \begin{subfigure}[b]{0.4\textwidth}
                \centering
                \includegraphics[scale=0.3]{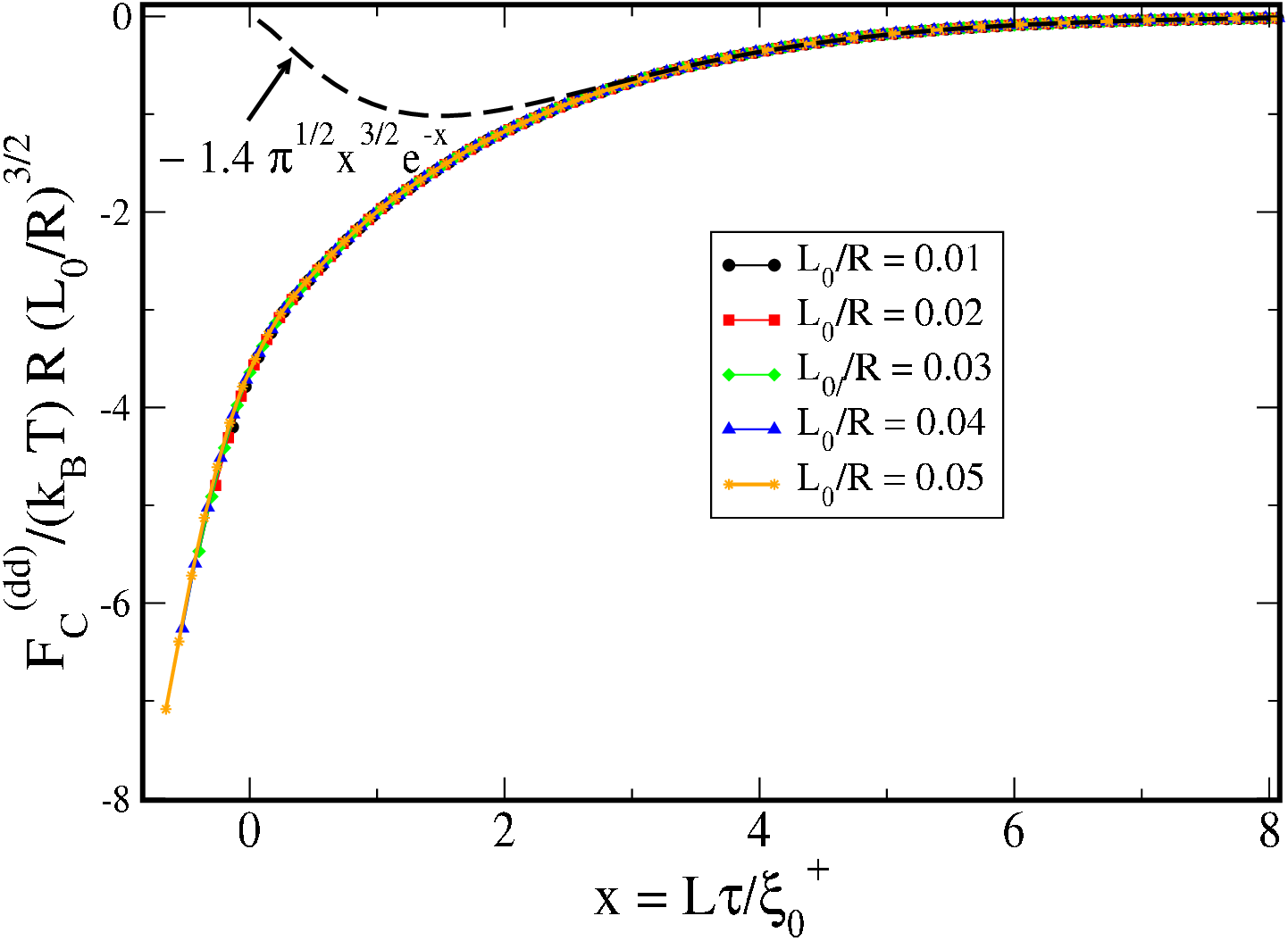}
                \caption{}

        \end{subfigure}%
        
		\vspace{0.85cm}        
        
        \begin{subfigure}[b]{0.4\textwidth}
                \centering
                \includegraphics[scale=0.3]{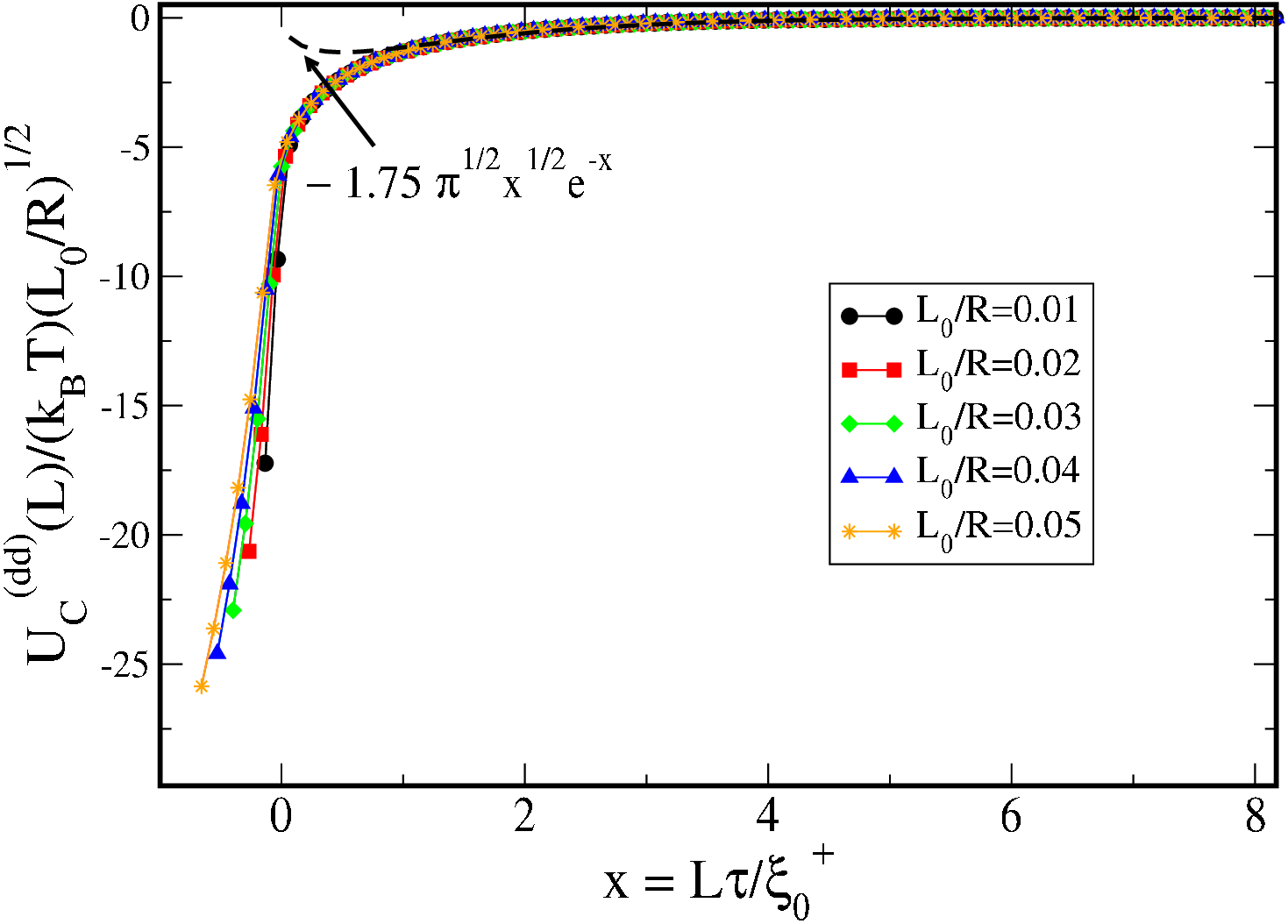}
                \caption{}
 
        \end{subfigure}
\caption{(Color online) The scaling function of the critical Casimir force (a) and of the critical Casimir potential (b)
the for two identical discs with radii $R=300, 500$ and 1000 as a function of the scaling variable
$x=L\tau/\xi_0^+$ at $z=20 000$ and $y = -17.8$. Dashed line  corresponds to $F_C^{(dd)}/(k_BT)$ and
$U_C^{(dd)}/(k_BT)$ calculated using the asymptotic behavior of the critical Casimir force for a $(+,+)$
strip (see the main text).}
\label{fig:10}
\end{figure}

Using the above defined variables $dS(\theta)$ and $L(\theta)$, and including the contribution from the lower
half of the disc (it means two integrals), we arrive at the following expression:

\begin{multline}
\label{finite}
\frac{F_C^{(dd)}}{k_BT}=2\frac{w}{L_0}\int_{0}^{\frac{\pi}{2}} \frac{\cos\theta}{(1+2w(1-\cos\theta))^2}\\
\times \vartheta (1+2w(1-\cos\theta))x,(1+2w(1-\cos\theta))y)d\theta,
\end{multline}
where $w=R/L_0$.
The potential $U_C^{(dd)}(L_0)$ associated with the Casimir force is given by:
\begin{equation}
\label{eq:def_pot}
\frac{U_C^{(dd)}}{k_BT}=-\int_{L_0}^{\infty}\frac{F_C^{(dd)}}{k_BT}(L)dL.
\end{equation}
We have calculated the disc-disc critical Casimir force and its potential numerically using the above formulae.   
Using  the known exact result for an asymptotic behavior of the Casimir scaling function for $2D$ Ising $(+,+)$  strip  at $y=0$ \cite{AM}, we determine
the asymptotic decay for $F_C^{(dd)}/k_BT$ and $U_C^{(dd)}/k_BT$ analytically using the following simplification of the integral
in (\ref{finite}).
In the Derjaguin limit $w=R/L_0 \gg 1$, this integral due to the denominator is dominated by  contributions 
corresponding to small values of the angle $\theta$, so the cosine function can be expanded to the second order
$\cos \theta \approx 1 - 1/2 \theta^2$ giving
\begin{equation}
\label{finite2}
\frac{F_C^{(dd)}}{k_BT}=2\frac{w}{L_0}\int_{0}^{\frac{\pi}{2}} \frac{1-\frac{1}{2}\theta^2}{(1+w\theta^2)^2}
\vartheta((1+w\theta^2)x)d\theta,
\end{equation}
Introducing the variable $l=w^{1/2}\theta$ and taking into account that for $w \rightarrow \infty$
the upper limit of the integral $w^{1/2} \pi/2$ goes to infinity independently of $\theta$, the integral can be
extended  to infinity taking the following form
\begin{multline}
\label{finite3}
\frac{F_C^{(dd)}}{k_BT}=2\frac{w^{3/2}}{R}\int_{0}^{\infty} \frac{1}{(1+l^2)^2}\vartheta((1+l^2)x)dl\\
-\frac{w^{1/2}}{R}\int_{0}^{\infty} \frac{l^2}{(1+l^2)^2}\vartheta((1+l^2)x)dl
\end{multline}
Since $l \sim \theta$, the value $l^2$ is very small, so that the second integral is subdominant and can be neglected.
As for the two-dimensional Ising strip (+,+) the finite-size scaling functions of the Casimir force has
the form $\vartheta(x)=-x^{2}e^{-x}$   for  $x\to +\infty$ \cite{AM}, one finds the following form of the leading integral
\begin{equation}
\label{finite4}
\frac{F_C^{(dd)}}{k_BT}=-\frac{w^{3/2}}{R} \pi^{1/2} x^{3/2} e^{-x} \qquad x\to +\infty.
\end{equation}
Accordingly to the above formula and from Eq.~(\ref{finite4}) we obtain the following asymptotic form for the potential
\begin{equation}
\label{eq:def_pot2}
\frac{U_C^{(dd)}}{k_BT}=-\pi^{1/2} w^{1/2} x^{1/2} e^{-x} \qquad x\to +\infty.
\end{equation}

Based on the general scaling arguments, $U_C^{(dd)}$ between two discs of a radius $R$ should take the following scaling form:
\begin{multline}
\label{eq:scal_pot}
\frac{U_C^{(dd)}(L_0;T,h,R)}{k_BT}=\frac{R}{L_0}\Theta(w,x,y)\\
=\frac{R}{L_0}\Theta_I(w,x,y/x=sgn(\tau h)\xi_{\tau}/\xi_h).
\end{multline}
Accordingly, for $F_C^{(dd)}$ one expects
\begin{multline}
\label{eq:scal_force}
\frac{F_C^{(dd)}(L_0;T,h,R)}{k_BT}=\frac{R}{L_0^2}\hat\vartheta(w,x,y)\\
=\frac{R}{L_0^2}\hat\vartheta_I(w,x,y/x=sgn(\tau h)\xi_{\tau}/\xi_h).
\end{multline}
For $x=0$ and  $y=0$,   the conformal invariance predicts  $\Theta(w,0,0) = -\frac{\pi^2}{24} \left(w\right)^{-1/2}$ 
in the Derjaguin limit (see Eq.~(\ref{eq:confinv})). It follows that
$\hat\vartheta(w,0,0) = -\frac{\pi^2}{48} \left(w\right)^{-1/2}$. Guided by the behavior at the bulk critical point,
in Fig.~\ref{fig:09}(a),(b)  we have plotted $F_C^{(dd)}/(k_BT)$ obtained from Eq.~(\ref{finite}), and rescaled by 
$(w^{3/2}R^{-1})^{-1}$ and $U_C^{(dd)}/(k_BT)$ obtained from Eq.~(\ref{eq:def_pot}) and rescaled by $(w^{1/2})^{-1}$, respectively,
as a function of the scaling variable $x$ for $y=0$ and various small ratios of $R/L_0$.  We find
an excellent data collapse indicating that the leading dependence
of the scaling functions $\hat\vartheta(w,x,0)$ and $\Theta(w,x,0)$ on the variable $w$ is the same as that at $x=0$
for $w \ll 1$.

Due to high costs of numerical calculation, we were not able  to produce all data necessary to 
determine the scaling functions $\hat\vartheta$ and $\Theta$ for fixed value of $y$. As follows from Eq.~(\ref{finite}),
in order to keep the argument $y$ of $\vartheta$ fixed, the function $\vartheta$ has to be known at many values
of the second variable. Nevertheless, we can  used our data for a strip with the  fixed value of $y$.
Because of the equality
\begin{widetext}
\begin{equation}
\label{fixed_y}
 \frac{F_C^{(dd)}}{k_BT}=\int_{\theta=0}^{\theta_M=\frac{\pi}{2}} \frac{dS(\theta)}{L^2(\theta)}\hat\vartheta(x(\theta),y(\theta))=
 \int_{\theta=0}^{\theta_M=\frac{\pi}{2}} \frac{dS(\theta)}{L^2(\theta)}\hat\vartheta_I\left((x(\theta),
 y(\theta)/x(\theta)=sgn(\tau h)\frac{\xi_{\tau}}{\xi_h}\right),
\end{equation}
\end{widetext}
this procedure determines the  Casimir force and the Casimir potential scaling function along the thermodynamic 
path $sgn(\tau h) (\xi_{\tau}/\xi_h)= constant$.

Results for $y=-17.8$ obtained using Eq.~(\ref{fixed_y}) are shown in Fig.~\ref{fig:10}(a),(b). There, 
we have plotted the  critical Casimir force and its potential, respectively, rescaled as 
for $y=0$ and  as a function of $x$. (For a strip, $y=-17.8$ corresponds to a path shown as the line
in Fig.~\ref{fig:01} passing closest to the pseudo critical capillary condensation point.) We can see that data
for $F_C^{(dd)}/(k_BT)$ collapse onto a master curve for the same values of the ratio $R/L_0$ as at $y=0$. For
the Casimir potential, deviations occur in the rapidly decaying part of the curve around $x=0$. These deviations
indicate that  corrections to the leading $w$ dependence of both scaling functions  become important.
Due to the  integration (\ref{eq:def_pot}), the deviations  are more pronounced  for the critical Casimir potential
than for the critical Casimir force.

Because of the limited number of data that we were able to obtain for a strip geometry, the Casimir scaling functions
for a disc-disc geometry have been extrapolated to large positive values of $x$. The extrapolation curves coincide with
the ones obtained by using the known asymptotic behavior of the Casimir force scaling function for $(+,+)$ Ising
strips with vanishing bulk field $y=0$ given by equations (Eqs.~(\ref{finite4}) and (\ref{eq:def_pot2}))
(dashed lines in Figs.~\ref{fig:09} and \ref{fig:10}). For $y\ne 0$, the asymptotic behavior of
$\hat\vartheta(x\to \infty, y=y_0)$ is not known, therefore using the same function as for $y=0$ is an approximation,
which works well but the amplitude differs from 1 and is equal to $1.6 \pm 0.2$.  
 
\section{Concluding remarks}
\label{sec:3}

In summary,  we have used the numerical DMRG  method in order to study 
the behavior of critical Casimir force $F_C$ along the near-critical  iso-fields  in 
 symmetric Ising  strips with strong surface fields. 
In order to reach the scaling regime and to determined the Casimir scaling 
functions, guided by the experience from our previous studies, we have considered strips
as wide as  $L=301-501$.

Contrary to the  case of $L=301$,  results for  wider strips  suffer 
from  deviations from the smooth curve in the vicinity
of the bulk critical point. The main reason lies in the accuracy of the extrapolation procedure
to the value of the
free energy for the infinite system.
Although the systems  up to $L = 700$ were analyzed  and the DMRG accuracy was very high (a number of states
kept $m = 64$), it has occurred to be not sufficient. This is not surprising, since
in the limit of  $L \to \infty$, when approaching the critical point, the correlation length diverges.

The most interesting result of our calculation is the  behavior of $F_C$ 
along the iso-fields which lie between the bulk coexistence
and the capillary condensation line. There, the magnitude 
of the critical Casimir
forces continuously increases  upon decreasing temperature. This increase is substantial -
 for the iso-field line which is the closest to the capillary condensation critical point, i.e.,
for $y=-17.8$, the force
becomes about 20 times stronger upon lowering the scaling variable $x$ from  the value -5 
to 5.
It may have an important consequences 
for experimental realizations, e.g.,  for  
lipid membranes with protein inclusions, because most probably the concentration of  components of the 
 membrane is not exactly at its critical value.

As an outlook, we would like  to calculate the scaling function of the disc-disc Casimir potential along
the thermodynamic paths which are accessible in experiments, i.e., changing the temperature at fixed $y$.

\section{Acknowledgements}
Numerical calculations were performed in WCSS Wroc\l aw (Poland, grant 82). We would like to thank dr Tomasz
Mas\l owski for his help in the numerical calculations. The author (M.Z.) is a scholar within Sub-measure 8.2.2
Regional Innovation Strategies, Measure 8.2 Transfer of knowledge, Priority VIII Regional human resources
 for the economy Human Capital Operational Programme co-financed by European Social Fund and state budget.


\begin{thebibliography}{99}
\bibitem{FdG} 
M. E. Fisher and P. G. de Gennes, C. R. Acad. Sci. Paris Ser. B {\bf 287}, 207 (1978).
%
\bibitem{krech:99:0} M. Krech, {\it Casimir Effect in Critical Systems} (World Scientific, Singapore, 1994); J. Phys.: Condens. Matter {\bf 11}, R391 (1999).
%
\bibitem{dantchev}  J. G. Brankov, D. M. Dantchev, and N. S. Tonchev, {\it The Theory of Critical Phenomena in Finite-Size Systems - Scaling and Quantum Effects} (World Scientific, Singapore, 2000).
%
\bibitem{gambassi} for a recent review see A. Gambassi, J. Phys.: Conf. Ser. {\bf 161}, 012037 (2009).
%
\bibitem{bechin}
O. Zvyagolskaya, A. J. Archer, and C. Bechinger, EPL {\bf 96}, 28005 (2011).
%
\bibitem{thomas1}   T. F. Mohry, A. Macio\l ek, and S. Dietrich, J. Chem. Phys. {\bf 136}, 224902 (2012).
%
\bibitem{schall} H. Guo, T. Narayanan, M. Sztuchi, P. Schall, and G. H. Wegdam, 
Phys. Rev. Lett. {\bf 100}, 188303 (2008).
%
\bibitem{schall1} V. D. Nguyen, S. Faber, Z. Hu, G. H. Wegdam and  P. Schall,  
Nature Commun. {\bf 4}, 1584 (2013).
%
\bibitem{thomas2}    T. F. Mohry, A. Macio\l ek, and S. Dietrich, J. Chem. Phys. {\bf 136}, 224903 (2012).
%
\bibitem{schall2}
S. J. Veen, O. Antoniuk, B. Weber, M. A. C. Potenza, S. Mazzoni, P. Schall, and G. H. Wegdam,
Phys. Rev. Lett. {\bf 109}, 248302 (2012).
%
\bibitem{Beysens-et:1985} 
D. Beysens and D. Est\`eve, Phys. Rev. Lett. {\bf 54}, 2123 (1985).
%
\bibitem{Beysens-et:1999} 
(a) D. Beysens, J.-M. Petit, T. Narayanan, A. Kumar, and M. L. Broide,
Ber. Bunsenges. Phys. Chem. {\bf 98}, 382 (1994);
(b) D. Beysens and T. Narayanan, J. Stat. Phys. {\bf 95}, 997 (1999).
%
\bibitem{Bonn-et:2009}
(a) D. Bonn, J. Otwinowski, S. Sacanna, H. Guo, G. Wegdam, and P. Schall,
Phys. Rev. Lett. {\bf 103}, 156101 (2009); 
(b) A. Gambassi and S. Dietrich, Phys. Rev. Lett. {\bf 105}, 059601 (2010);
(c) D. Bonn, G. Wegdam, and P. Schall, Phys. Rev. Lett. {\bf 105}, 059602 (2010).
%
\bibitem{Buzzaccaro-et:2010} 
(a) S. Buzzaccaro, J. Colombo, A. Parola, and R. Piazza, 
Phys. Rev. Lett. {\bf 105}, 198301 (2010); 
(b) R. Piazza, S. Buzzaccaro, A. Parola, and J. Colombo, 
J. Phys.: Condens. Matter {\bf 23}, 194114 (2011). 
%
\bibitem{dme1} A. Drzewi\'nski, A. Macio\l ek, and R. Evans,
Phys. Rev. Lett. {\bf 85 }, 3079 (2000); A. Macio\l ek, A. Drzewi\'nski, and R. Evans,
Phys. Rev. E  {\bf 64}, 056137 (2001).
%
\bibitem{dme2} A. Drzewi\'nski, A. Macio\l ek, and A. Ciach,
Phys. Rev. E {\bf 61 }, 5009 (2000); A. Macio\l ek, A. Drzewi\'nski, and A. Ciach,
Phys. Rev. E  {\bf 64}, 026123 (2001).
%
\bibitem{colloids1b}
F. Schlesener, A. Hanke, and S. Dietrich, 
J. Stat. Phys. {\bf 110}, 981 (2003).
%
\bibitem{onuki}
R. Okamoto and A. Onuki, J. Chem. Phys. {\bf 136}, 114704 (2012).
%
\bibitem{thesis}
T. F. Mohry, doctoral thesis, University of Stuttgart (2013).
%
\bibitem{Derjaguin:1934} 
B. Derjaguin, Kolloid Zeitschrift {\bf 69}, 155 (1934).
%
\bibitem{diehl} H. W. Diehl in {\it Phase Transitions and Critical Phenomena},
edited by C. Domb and J. L. Lebowitz (Academic, New York, 1986), Vol. 10,
p. 76.
%
\bibitem{diehl1} T. W. Burkhardt and H. W. Diehl, Phys Rev. B {\bf 50}, 3894 (1994).
%
\bibitem{DMRG} U.~Schollwoeck, Rev.~Mod.~Phys. {\bf 77}, 259 (2005); K.~Hallberg, Adv.~Phys. {\bf 55}, 477 (2006);
U.~Schollwoeck, Ann.~Phys.~(NY) {\bf 326}, 96 (2011).
%
\bibitem{Nishino} T.~Nishino, and K.~Okunishi, J.~Phys.~Soc.~Jpn. {\bf 65}, 89 (1996); K.~Ueda, R.~Kr\v{c}m\'{a}r,
A.~Gendiar, and T.~Nishino, J.~Phys.~Soc.~Jpn. {\bf 76}, 084004 (2007).
%
\bibitem{AD} E.~Carlon, A.~Drzewi\'{n}ski, and J.~Rogiers, Phys. Rev.{\bf B 58}, 5070 (1998); A.~Drzewi\'{n}ski,
A.~Macio\l ek and R.~Evans, Phys. Rev. Lett. {\bf 85}, 3079 (2000); A.~O.~Parry, A.~J.~Wood, E.~Carlon, and
A.~Drzewinski, Phys.~Rev.~Lett. {\bf 87}, 196103 (2001); A.~Drzewi\'{n}ski, A.~Macio\l ek, A.~Barasi\'{n}ski, and
S.~Dietrich, Phys.~Rev. {\bf E 79}, 041145 (2009).
%
\bibitem{LS}
D. Lingwood and K. Simons,
Science {\bf 327}, 46 (2010).
%
\bibitem{membranes1}  S. L. Veatch and S. L. Keller,
Biochim. Biophys. Acta, Mol. Cell Res. {\bf 1746}, 172 (2005).

\bibitem{membranes} 
A. R. Honerkamp-Smith, P. Cicuta, M. D.  Collins, S. L. Veatch, M. den Nijs,  
M. Schick, and S. L. Keller, Biophys. J. {\bf 95}, 236 (2008);
M. C. Heinrich, I.  Levental, H. Gelman,P. A. Janmey, and T.  Baumgart, J. Phys. Chem B {\bf 112}, 8063, 2008.
%
\bibitem{sehtna}
B. B. Machta, S. Papanikolaou, J. P. Sethna, and S. L. Veatch,
Biophys. J. {\bf 100}, 1668 (2011).
%
\bibitem{vmd}
O. Vasilyev, A. Macio\l ek, and S. Dietrich,
Phys. Rev. E {84}, 041605 (2011). 
%
\bibitem{TF} H. B.  Tarko and M. E. Fischer, Phys. Rev. B, {\bf 11}, 1217 (1975).

%
\bibitem{DMC} A. Drzewi\'nski, A. Macio\l ek, and A. Ciach, Phys. Rev E, {\bf 61}, 5009, (2000).
%
\bibitem{Onsager} L.~Onsager, Phys.~Rev. {\bf 65}, 117 (1944); B.~Kaufman, and L.~Onsager,
Phys.~Rev. {\bf 76}, 1244 (1949).
%
\bibitem{BS} J.~Stoer, and Bulirsch, {\em Introduction to Numerical Analysis}, (New York: Springer-Verlag, 1980).
%
\bibitem{FSS} M. N. Barber in  {\it Phase Transitions and Critical Phenomena}, edited by C. Domb and J. L. Lebowitz, Vol.~{\bf 8}
(Academic, London, 1983), p. 145.
%
\bibitem{Bob} R.~Evans, and J.~Stecki, Phys.~Rev. B {\bf 49}, 8842 (1994).
%
\bibitem{BE}
T. W. Burkhardt and E. Eisenriegler, Phys. Rev. Lett. {\bf 74}, 3189 (1995).
%
\bibitem{MS}
 B. B. Machta,  J. P. Sethna, and S. L. Veatch, Phys. Rev. Lett. {\bf 109}, 138101 (2012).
%
\bibitem{Hanke}
A. Hanke, F.  Schlesener, E. Eisenriegler, and S. Dietrich, Phys. Rev. Lett. {\bf 81}, 1885 (1995);
S. Kondrat, L. Harnau,  and S. Dietrich, J. Chem. Phys. {\bf 131}, 204902 (2009);
F. Schlesener, A. Hanke, and S. Dietrich, J. Stat. Phys. {\bf 110}, 981 (2003).
M. Tr\"{o}ndle,  S. Kondrat,  A. Gambassi, L. Harnau, and S. Dietrich, J. Chem. Phys. {\bf 133}, 074702 (2010);
M. Hasenbusch,  Phys. Rev. E {\bf 87}, 022130 (2013). 

\bibitem{AM} D. B. Abraham and A. Macio\l ek, EPL {\bf 101}, 20006 (2013).

\end{thebibliography}
\end{document}